\def\be{\begin{equation}}
\def\ee{\end{equation}}
\def\ba{\begin{array}}
\def\ea{\end{array}}

\newcommand{\lab}{\label}

\documentclass[aps,pra,showpacs,showkeys,preprint]{revtex4}
\usepackage{amsmath}
\usepackage{amssymb}
\usepackage{graphicx}
\begin{document}

\title{Lower Bounds of Concurrence for Tripartite Quantum Systems}

\author{Xiu-Hong Gao$^{a}$}
\author{Shao-Ming Fei$^{a,b}$}
\author{Ke Wu$^{a}$}

\affiliation{$~^{a}$ Department of Mathematics, Capital Normal University, Beijing
100037, China\\
$~^{b}$ Max Planck Institute for Mathematics in the Sciences, D-04103 Leipzig,
Germany}

\begin{abstract}
We derive an analytical lower bound for the concurrence of
tripartite quantum mixed states. A functional relation is
established relating concurrence and the generalized partial
transpositions.
\end{abstract}

\maketitle

\bigskip
\medskip

As the key physical resources in quantum information processing
and quantum computation \cite{nielsen}, the quantum entangled
states have been investigated with a great deal of effort in the past
years [2-14]. So far for generic mixed states only partial
solutions are known on detection and quantification of entanglement in an
operational way. Concurrence is one of the well defined
quantitative measures of entanglement. For two-qubit case
another measure, entanglement of formation \cite{BDSW,Horo-Bruss-Plenioreviews} is a
monotonically increasing function of concurrence and
an elegant formula of concurrence was derived analytically by
Wootters in \cite{Wootters98}, which
plays an essential role in describing quantum phase
transition in various interacting quantum many-body systems
\cite{Osterloh02-Wu04} and may affect macroscopic properties of
solids significantly \cite{Ghosh2003}.
What is more, it can be experimentally measured \cite{buchleitner}.

Nevertheless, calculation of the concurrence is a formidable task for
higher dimensional case. Therefore some nice
algorithms and progresses have been concentrated on
possible lower bounds of the concurrence for qubit-qudit systems
\cite{chenp02-Gerjuoy03,Lozinski03} and for
bipartite systems in arbitrary dimensions
\cite{Audenaert01,Mintert04-MintertPhD} but involving
numerical optimization over a large number of
free parameters. In \cite{Chen-Albeverio-Fei}
an analytical lower bound of concurrence for any dimensional mixed
bipartite quantum states has been presented, which is further shown to be exact
for some special classes of states and detects many bound entangled states.

Although the lower bound for entanglement of formation can be
similarly investigated for bipartite case
\cite{Chen-Albeverio-Fei1}, for tripartite case the entanglement of
formation is not yet well defined. In contract, the concurrence for
tripartite states is well defined. In this paper we consider the
lower bound of concurrence for tripartite states, by exploring the
connection between the generalized partial transposition (GPT)
criterion and concurrence.

Let ${\mathcal H}_A$, ${\mathcal H}_B$ and ${\mathcal H}_C$ be three
finite dimensional Hilbert spaces associated with the subsystems
$A$, $B$ and $C$, with dimensions $dim~A=m$, $dim~B=n$ and $dim~C=p$.
The concurrence for a general pure tripartite state ${\left\vert \psi \right\rangle }
\in {\mathcal H}_A \otimes {\mathcal H}_B\otimes {\mathcal H}_C$
is defined by
\begin{equation}\label{cpure}
C({\left\vert \psi \right\rangle })=\sqrt{3-\mbox{Tr}(\rho _{A}^{2}
+\rho_{B}^{2}+\rho _{C}^{2})},
\end{equation}%
where the reduced density matrix
$\rho_{A}$ (resp. $\rho _{B}$, $\rho _{C}$) is obtained by tracing over the subsystems
$B$ and $C$ (resp. $A$ and $C$, $A$ and $B$). The concurrence for a tripartite mixed state $\rho$
is defined by the convex roof,
\begin{equation}\label{e1}
C(\rho )\equiv \min_{\{p_{i},|\psi _{i}\rangle
\}}\sum_{i}p_{i}C({\left\vert \psi _{i}\right\rangle }),
\end{equation}
for all possible ensemble realizations $\rho =\sum_{i}p_{i}|\psi
_{i}\rangle\langle \psi _{i}|$, where ${\left\vert \psi_i \right\rangle }
\in {\mathcal H}_A \otimes {\mathcal H}_B\otimes {\mathcal H}_C$,
$p_{i}\geq 0$ and $\sum_{i}p_{i}=1$. For any pure product state ${\left\vert \psi
\right\rangle }$, $C({\left\vert \psi \right\rangle })$ vanishes
according to the definition. Consequently, if a state $\rho $ is
\emph{separable}, then $C(\rho )=0$.

To get a lower bound of (\ref{e1}), we relate directly the
concurrence to the generalized partial transposition separability
criterion. We first recall some notations used in various matrix
operations \cite{Albeverio-Chen-Fei, hornt1,hornt}.

A generic matrix $M$ can be always written as $M={\displaystyle\sum\limits_{i,j}}a_{ij}\left\langle
j\right\vert \otimes \left\vert i\right\rangle $, where $\left\vert
i\right\rangle ,\left\vert j\right\rangle $ are vectors of a suitably selected normalized
real orthogonal basis. We define the operations $\mathcal{T}_{r}$ (resp. $\mathcal{T}_{c}$)
to be the row transposition (resp. column transposition) of $M$
which transposes the second (resp. first) vector in the above tensor product
expression of $M$:
\begin{equation}
\mathcal{T}_{r}(M)={\displaystyle\sum\limits_{i,j}}
a_{ij}\left\langle j\right\vert \otimes \left\langle i\right\vert,~~~~
\mathcal{T}_{c}(M)={\displaystyle\sum\limits_{i,j}}
a_{ij}\left\vert j\right\rangle \otimes \left\vert i\right\rangle.
\end{equation}
It is easily verified that
$\mathcal{T}_{c}\mathcal{T}_{r}(M)=\mathcal{T}_{r}\mathcal{T}_{c}(M)=M^{t}$,
where $t$ denotes matrix transposition.

We further define ${\mathcal{T}}_{r_{k}}$ (resp. ${\mathcal{T}}_{c_{k}}$)
($k=A,B,C,AB,BC,AC$) to be the row (resp. column) transpositions with respect to
the subsystems $k$.
Set $\mathcal{T}_{\{x_{1},x_{2},...\}}\equiv\mathcal{T}_{x_{1}}\mathcal{T}_{x_{2}}...$
for $x_{1},~x_{2}\subset \Gamma\equiv
\{r_{A}, c_{A}, r_{B}, c_{B}, r_{C}, c_{C},  r_{AB}, r_{AC}, r_{BC}, c_{AB}, c_{AC}, c_{BC}\}$.
We consider the generalized partial
transposition operations on a tripartite density matrix given by
${\mathcal{T}}_{\mathcal{ Y}}$,
where ${\mathcal{T}}_{\mathcal{Y}}$ stands for all partial
transpositions contained in ${\mathcal{Y}}$ which is a subset of
$\Gamma$. The GPT criterion says that if a tripartite $m\times n
\times p$ density matrix is separable, then the trace norm
$||\rho^{{\mathcal{T}}_{\mathcal{Y}}}||\leq 1$, where
$\rho^{{\mathcal{T}}_{\mathcal{Y}}}={{\mathcal{T}}_{\mathcal{Y}}}(\rho)$,
for instance $\rho^{{\mathcal{T}}_{\{c_A,r_B,r_C\}}}\equiv
{\mathcal{T}}_{\{c_A\}}{\mathcal{T}}_{\{r_B\}}{\mathcal{T}}_{\{r_C\}}(\rho)$
and so on. In the following we discuss three classes of
${\mathcal{Y}}$:

I: ${\mathcal{Y}}_i=\{c_{k},r_k\}$, where $i=1,2,3$ for $k=A,B,C$
respectively;

II: ${\mathcal{Y}}_4=\{c_{A},r_{BC}\},\,{\mathcal{Y}}_5=\{c_{AB},r_{C}\},\,
{\mathcal{Y}}_6=\{c_{AC},r_{B}\}$;

III: ${\mathcal{Y}}_7=\{c_{A},r_{B}\},\,
{\mathcal{Y}}_8=\{c_{A},r_{C}\},\,{\mathcal{Y}}_9=\{c_{B},r_{C}\}$.

It is verified that
$\rho^{{\mathcal{T}}_{{\mathcal{Y}}_i}}=\rho^{T_k}$, where $k=A,B,C$ with respect to
$i=1,2,3$, ${T_k}$ stands for partial transposition with respect to
the subsystem $k$. Hence the operations in class I correspond to the partial
transpositions of $\rho$. While the operations in class III correspond to the
realignments of a tripartite state $\rho$ \cite{Albeverio-Chen-Fei,chen-wu}.

We first study the relation between GPT and the concurrence for
three qubits ($m=n=p=2$). A three-qubit state $|\Psi\rangle$ can be
written in terms of the generalized Schmidt decomposition
\cite{acin},
\begin{eqnarray}
|\Psi\rangle &=& \lambda_0 |000\rangle+\lambda_1 e^{i\psi}
|100\rangle+\lambda_2 |101\rangle +\lambda_3 |110\rangle+\lambda_4
|111\rangle \label{geschmidt}
\end{eqnarray}
with normalization condition
$\lambda_i \geq 0,\ \ 0 \leq \psi \leq \pi$, where $\sum_i \mu_i=1$, $\mu_i \equiv\lambda_i^2$.

Defining $\Delta \equiv |\lambda_1 \lambda_4 e^{i \psi}-\lambda_2
\lambda_3|^2$, we have, for $\rho=|\Psi\rangle\langle\Psi|$
$$\ba{l}
Tr\rho_A^2=1-2\mu_0(1-\mu_0-\mu_1),\\[3mm]
Tr\rho_B^2=1-2\mu_0(1-\mu_0-\mu_1-\mu_2)-2\Delta,\\[3mm]
Tr\rho_C^2=1-2\mu_0(1-\mu_0-\mu_1-\mu_3)-2\Delta.
\ea
$$
Therefore
\begin{equation}
C^2(\rho)=2\mu_0(3-3\mu_0-3\mu_1-\mu_2-\mu_3)+4\Delta,
\label{concurrence}
\end{equation}
which varies smoothly from 0, for pure product states, to $\frac{3}{2}$ for
maximally entangled pure states.

On the other hand, we have
$$\rho^{{\mathcal{T}}_{{\mathcal{Y}}_1}} = \left(\ba{cccccccc}
\mu_0& 0& 0& 0& \lambda_0\lambda_1 e^{i\psi}& 0& 0&0\\
0& 0& 0& 0& \lambda_0\lambda_2& 0& 0& 0\\
0& 0& 0& 0& \lambda_0\lambda_3& 0& 0& 0\\
0& 0& 0& 0& \lambda_0\lambda_4& 0&0& 0\\
\lambda_0\lambda_1 e^{-i\psi}& \lambda_0\lambda_2&
\lambda_0\lambda_3& \lambda_0\lambda_4& \mu_1& \lambda_2\lambda_1
e^{i\psi}&
\lambda_3\lambda_1 e^{i\psi}& \lambda_4\lambda_1 e^{i\psi}\\
0& 0& 0&0& \lambda_1\lambda_2 e^{-i\psi}& \mu_2& \lambda_2\lambda_3& \lambda_2\lambda_4\\
0& 0& 0& 0& \lambda_3\lambda_1 e^{-i\psi}& \lambda_3\lambda_2& \mu_3&  \lambda_3\lambda_4\\
0& 0& 0& 0& \lambda_4\lambda_1 e^{-i\psi}& \lambda_4\lambda_2&
\lambda_4\lambda_3& \mu_4
\ea\right).
$$
As
$\rho^{{\mathcal{T}}_{{\mathcal{Y}}_1}}={\rho^{{\mathcal{T}}_{{\mathcal{Y}}_1}}}^\dag$,
the square root of the eigenvalues of
$\rho^{{\mathcal{T}}_{{\mathcal{Y}}_1}}{\rho^{{\mathcal{T}}_{{\mathcal{Y}}_1}}}^\dag$
is the absolute value of the eigenvalues of
$\rho^{{\mathcal{T}}_{{\mathcal{Y}}_1}}$: $\{0, 0, 0, 0,
\pm\sqrt{\mu_0(\mu_2+\mu_3+\mu_4)},
    \frac{1}{2}(1\pm\sqrt{1-4\mu_0(\mu_2+\mu_3+\mu_4)})\}$.
Therefore the norm of $\rho^{{\mathcal{T}}_{{\mathcal{Y}}_1}}$ is
given by
\begin{equation}
||\rho^{{\mathcal{T}}_{{\mathcal{Y}}_1}}|| =
1+2\sqrt{\mu_0(\mu_2+\mu_3+\mu_4)}. \label{pta}
\end{equation}
Similarly we have
\begin{eqnarray}
||\rho^{{\mathcal{T}}_{{\mathcal{Y}}_2}}|| &=&1+2 \sqrt{\Delta+\mu_0(\mu_3+\mu_4)}, \label{ptb}\\
||\rho^{{\mathcal{T}}_{{\mathcal{Y}}_3}}|| &=&1+2
\sqrt{\Delta+\mu_0(\mu_2+\mu_4)}. \label{ptc}
\end{eqnarray}

A lower bound for the concurrence of three-qubit states is given by
the following theorem.

 {\sf [Theorem 1].} For any three-qubit mixed
quantum state $\rho$, the concurrence $C(\rho )$ satisfies \be
C(\rho )\geq max\left\{\Vert
\rho^{{\mathcal{T}}_{{\mathcal{Y}}_i}}\Vert-1,~
\frac{1}{\sqrt{2}}(\Vert\rho^{{\mathcal{T}}_{{\mathcal{Y}}_j}}\Vert-1)
\right\}, \label{maintheorem1} \ee where $i=1,2,3$, $j=4,5,6$.

{\sf [Proof].} Let us assume that one
has already found an optimal decomposition $\sum_{i}p_{i}\rho ^{i}$ for $%
\rho $ to achieve the infimum of $C(\rho )$, where $\rho ^{i}$ are
pure state density matrices. Then $C(\rho )=\sum_{i}p_{i}C(\rho
^{i})$ by definition. Noticing that $\Vert
\rho^{{\mathcal{T}}_{{\mathcal{Y}}_j}}\Vert \leq \sum_{i}p_{i}\Vert
(\rho^i)^{{\mathcal{T}}_{{\mathcal{Y}}_j}}\Vert $, for all possible
$j$, due to the convex property of the trace norm, one only needs to
show $C(\rho ^{i})\geq (\Vert
(\rho^i)^{{\mathcal{T}}_{{\mathcal{Y}}_j}}\Vert -1)$ for $j=1,2,3$
and $C(\rho ^{i})\geq \frac{1}{\sqrt{2}}(\Vert
(\rho^i)^{{\mathcal{T}}_{{\mathcal{Y}}_j}}\Vert -1)$, for $j=4,5,6$.

For a pure state $\rho ^{i}$, from Eqs.~(\ref{concurrence}) , (\ref{pta}),
(\ref{ptb}) and (\ref{ptc}), we have
$$
\ba{l} C^2(\rho^i)-(\Vert
(\rho^i)^{{\mathcal{T}}_{{\mathcal{Y}}_1}}\Vert-1)^2
=2\mu_0\mu_4+4\Delta\geq 0,\\[3mm]
C^2(\rho^i)-(\Vert
(\rho^i)^{{\mathcal{T}}_{{\mathcal{Y}}_2}}\Vert-1)^2
=4\mu_0\mu_2+2\mu_0\mu_4 \geq 0 \ea
$$
and
$$
C^2(\rho^i)-(\Vert
(\rho^i)^{{\mathcal{T}}_{{\mathcal{Y}}_3}}\Vert-1)^2
=4\mu_0\mu_3+2\mu_0\mu_4 \geq 0.  \label{C-N}
$$
That is $C(\rho ^{i})\geq (\Vert
(\rho^i)^{{\mathcal{T}}_{{\mathcal{Y}}_j}}\Vert -1)$ for $j=1,2,3$.

For a pure state $\rho^i$, we consider it as a
$2\otimes 4$,  or $4\otimes 2$ bipartite state, respectively. From the results for
bipartite systems \cite{Chen-Albeverio-Fei}, we have
$$\ba{l}
1-Tr((\rho^i_A)^2)\geq
\displaystyle\frac{1}{2}(||(\rho^i)^{{\mathcal{T}}_{\{c_A,r_{BC}\}}}||-1)^2,\\[3mm]
1-Tr((\rho^i_B)^2)\geq
\displaystyle\frac{1}{2}(||(\rho^i)^{{\mathcal{T}}_{\{c_{AC},r_B\}}}||-1)^2,\\[3mm]
1-Tr((\rho^i_C)^2)\geq
\displaystyle\frac{1}{2}(||(\rho^i)^{{\mathcal{T}}_{\{c_{AB},r_C\}}}||-1)^2.
\ea
$$
Therefore
\be
\begin{array}{ll}
C(\rho^i)=\sqrt{3-Tr((\rho^i_A)^2)-Tr((\rho^i_B)^2)-Tr((\rho^i_C)^2)}\geq \\[4mm]
\displaystyle\frac{1}{\sqrt{2}}max\left\{(||(\rho^i)^{{\mathcal{T}}_{\{c_A,r_{BC}\}}}||-1),
(||(\rho^i)^{{\mathcal{T}}_{\{c_{AC},r_B\}}}||-1),
(||(\rho^i)^{{\mathcal{T}}_{\{c_{AB},r_C\}}}||-1)\right\},
\label{pure-c(p)-cab-rc}
\end{array}
\ee
i.e.
$$
C(\rho^i)\geq \frac{1}{\sqrt{2}} max
\left\{||(\rho^i)^{{\mathcal{T}}_{{\mathcal{Y}}_j}}||-1\right\},~~~j=4,5,6,
$$
which ends the proof. \hfill$\Box$

As an example, let us consider the D\"{u}r-Cirac -Tarrach states
\cite{dur-cirac-tarrach}:
\begin{equation}\label{e32}
\rho=\sum_{\sigma=\pm}
\lambda_0^{\sigma}|\Psi_0^{\sigma}\rangle\langle\Psi_0^{\sigma}|+\sum_{j=1}^3
\lambda_j(|\Psi_j^{+}\rangle\langle\Psi_j^{+}|+|\Psi_j^{-}\rangle\langle\Psi_j^{-}|),
\end{equation}
where the orthonormal Greenberger-Horne-Zeilinger (GHZ)-basis
$$|\Psi_j^{\pm}\rangle\equiv\frac{1}{\sqrt{2}}
(|j\rangle_{AB}|0\rangle_C\pm|(3-j)\rangle_{AB}|1\rangle_C),
$$
$|j\rangle_{AB}\equiv|j_1\rangle_{A}|j_2\rangle_{B}$ with $j=j_1j_2$
in binary notation. For example,
$|\Psi_0^{\pm}\rangle\equiv\frac{1}{\sqrt{2}}(|000\rangle\pm|111\rangle)$
is the standard GHZ states.

A direct calculation gives rise to
$||\rho^{{\mathcal{T}}_{{\mathcal{Y}}_1}}||=\frac{4}{3}$,
$||\rho^{{\mathcal{T}}_{{\mathcal{Y}}_2}}||=||\rho^{{\mathcal{T}}_{{\mathcal{Y}}_3}}||=1$,
and $||\rho^{{\mathcal{T}}_{{\mathcal{Y}}_j}}||=0.8727$ for
$j=4,5,6$. Therefore, $C(\rho)\geq \frac{1}{3}$ according to theorem
1 for
$\lambda_0^{+}=\frac{1}{3};\lambda_1=\lambda_3=\frac{1}{6};\lambda_0^{-}=\lambda_2=0$.
This shows that the state is entangled, which is also a conclusion
implied by \cite{dur-cirac-tarrach,yu-song}.

We have obtained lower bounds of the concurrence in terms of the generalized
partial transposition. Similar to the bipartite case, it is
also possible to find lower bounds of the concurrence in terms of
the realignment operations, described in class III, which correspond to
the realignments of the density matrix $\rho$
on $A,B$; $A,C$ and $B,C$ subsystems, while leaving the remaining $C$; $B$
and $A$ subsystems unchanged. For instance, with respect to the operation ${{\mathcal{Y}}_7}$,
$\rho^{{\mathcal{T}}_{\{c_A,r_B\}}}$ implies
$\rho^{{\mathcal{T}}_{\{c_A,r_B\}}}_{ijm,kln}=\rho _{ikm,jln}$, where
the indices $i(k,m)$ and $j(l,n)$ are viewed as the row and column indices for the
subsystem $A(B,C)$ respectively.

Let us consider a special-type of three-qubit states
by setting $\lambda_i=0$, $i=1,2,3$ in (\ref{geschmidt}),
\begin{eqnarray}
|\Phi\rangle &=&\lambda_0 |000\rangle+\lambda_4 |111\rangle
\label{specialtype}
\end{eqnarray}
with normalization condition
$\lambda_0,~\lambda_4 \geq 0$, $\lambda_0^2+\lambda_4^2=1$.
We get, for $\rho_0=|\Phi\rangle\langle\Phi|$,
$$
\rho^{{\mathcal{T}}_{\{c_A,r_B\}}}_0=\lambda_0^2 |000\rangle\langle000|+
\lambda_0\lambda_4|010\rangle\langle011|+
\lambda_0\lambda_4|101\rangle\langle100|+
\lambda_4^2|111\rangle\langle111|.
$$
Hence the sum of its singular values is
$||\rho^{{\mathcal{T}}_{\{c_A,r_B\}}}_0||=1+2\lambda_0\lambda_4$.
Similarly, we have
$||\rho^{{\mathcal{T}}_{\{c_A,r_C\}}}_0||=||\rho^{{\mathcal{T}}_{\{c_B,r_C\}}}_0||=1+2\lambda_0\lambda_4$.

From Eqs.~(\ref{concurrence}), (\ref{pta}), (\ref{ptb}) ,
(\ref{ptc}) and  direct calculations we have at last
$C(\rho_0)=\sqrt{6\mu_0\mu_4}$,
$||\rho^{{\mathcal{T}}_{{{\mathcal{Y}}_j}}}_0||=2\lambda_0\lambda_4+1$, $j=1,...,9$.
By using the procedure in proving Theorem 1, we arrive at:

{\sf [Corollary].} For any three-qubit mixed state with
decomposition $\rho=\sum_i p_i |\Psi_i\rangle \langle \Psi_i|$, if
$|\Psi_i\rangle$ can be written in the form (\ref{specialtype}) for
any $i$, then the concurrence $C(\rho )$ satisfies
\begin{equation}
C(\rho )\geq max\{||\rho^{{\mathcal{T}}_{{\mathcal{Y}}_j}}||\}-1,~~~j=1,...,9.
\label{corollary}
\end{equation}

We remark that once a density matrix has a decomposition with all the
pure states of the form (\ref{specialtype}), then its all other possible
decompositions will also have the form (\ref{specialtype}),
since other decompositions can be obtained from the unitarily linear combinations
of this decomposition, and any linear combinations of the type (\ref{specialtype})
still have the form (\ref{specialtype}).

Although for general three-qubit states we do not have an analytical relation between
concurrence and the realignment operations (Class III),
the numerical computations imply that
$C(\rho)\geq max\{||\rho^{{{\mathcal{T}}_{{\mathcal{Y}}_j}}}||\}-1$, $j=7,8,9$, is still valid.
We chose $10^6$ random vectors
$(\lambda_0,\lambda_1,\lambda_2,\lambda_3,\lambda_4,\psi )$
for state (\ref{geschmidt}), calculated $C(\rho)$
and $||\rho^{{\mathcal{T}}_{{\mathcal{Y}}_j}}||-1$,
$j=7,8,9$. All results agree with the inequality in Corollary.
From the proof of Theorem 1, it implies
that the inequality would be also correct for mixed states.

Generalizing the results of Theorem 1 to arbitrary dimensional tripartite quantum
states, we have the following lower bounds:

{\bf [Theorem 2].} For any $m\otimes n\otimes p$ $(m\leq n,p)$
tripartite mixed quantum state $\rho$, the concurrence $C(\rho)$
satisfies
\be
C(\rho)\geq max\left\{\sqrt{\frac{1}{m(m-1)}}(||\rho^{{\mathcal{T}}_{{\mathcal{Y}}_a}}||-1),
\sqrt{\frac{1}{q(q-1)}}(||\rho^{{\mathcal{T}}_{{\mathcal{Y}}_b}}||-1),
\sqrt{\frac{1}{r(r-1)}}(||\rho^{{\mathcal{T}}_{{\mathcal{Y}}_c}}||-1)\right\},\lab{c-ppt}
\ee
where $q=min{(n,mp)}$ and $r=min{(p,mn)}$,
${\mathcal{Y}}_a={\mathcal{Y}}_1$ or ${\mathcal{Y}}_4$,
${\mathcal{Y}}_b={\mathcal{Y}}_2$ or ${\mathcal{Y}}_6$,
${\mathcal{Y}}_c={\mathcal{Y}}_3$ or ${\mathcal{Y}}_5$.

{\bf [Proof]}. Let us assume that one has already found an optimal
decomposition $\sum_i p_i \rho^i$ for $\rho$ to achieve the infimum
of $C(\rho)$, where $\rho^i$ are pure state density matrices. Then
$C(\rho)=\sum_i p_i C(\rho^i)$ by definition. Noticing that
$||\rho^{{\mathcal{T}}_{{\mathcal{Y}}_k}}||\leq \sum_i p_i
||(\rho^i)^{{\mathcal{T}}_{{\mathcal{Y}}_k}}||$, $k=a,b,c$, due to
the convex property of the trace norm, one only needs to show
$C(\rho^i)\geq
\sqrt{\frac{1}{j(j-1)}}(||(\rho^i)^{{\mathcal{T}}_{{\mathcal{Y}}_k}}||-1)$,
where $j=m$, $q$, $r$ for $k=a$, $b$, $c$, respectively. For a pure
state $\rho^i$, we consider it as a $m\otimes np$, $n\otimes mp$ or
$mn\otimes p$ bipartite state, respectively. From the result of
\cite{Chen-Albeverio-Fei}, we obtain
$$\begin{array}{l}
1-Tr((\rho^i_A)^2)\geq \displaystyle\frac{1}{m(m-1)}(||(\rho^i)^{{\mathcal{T}}_{{\mathcal{Y}}_a}}||-1)^2,\\[4mm]
1-Tr((\rho^i_B)^2)\geq \displaystyle\frac{1}{q(q-1)}(||(\rho^i)^{{\mathcal{T}}_{{\mathcal{Y}}_b}}||-1)^2,\\[4mm]
1-Tr((\rho^i_C)^2)\geq
\displaystyle\frac{1}{r(r-1)}(||(\rho^i)^{{\mathcal{T}}_{{\mathcal{Y}}_c}}||-1)^2.
\end{array}
$$
Therefore from the definition of $C(\rho^i)$,
\be
\begin{array}{ll}
\sqrt{3-Tr((\rho^i_A)^2)-Tr((\rho^i_B)^2)-Tr((\rho^i_C)^2)}\geq \\[4mm]
max\left\{\sqrt{\displaystyle\frac{1}{m(m-1)}}(||\rho^{{\mathcal{T}}_{{\mathcal{Y}}_a}}||-1),
\sqrt{\displaystyle\frac{1}{q(q-1)}}(||\rho^{{\mathcal{T}}_{{\mathcal{Y}}_b}}||-1),
\sqrt{\displaystyle\frac{1}{r(r-1)}}(||\rho^{{\mathcal{T}}_{{\mathcal{Y}}_c}}||-1)\right\}.
\label{pure-c(p)}
\end{array}
\ee
\hfill$\Box$

In summary, by making a novel connection with the
generalized partial transpositions, we have provided an entirely analytical formula for
lower bound of concurrence for tripartite systems.
One only needs to calculate the trace norm of certain matrices,
which avoids complicated optimization procedure over a large number
of free parameters in numerical approaches.
The results could be used to indicate possible
quantum phase transitions in condensed matter systems, and to
analyze finite size or scaling behavior of entanglement in
various interacting quantum many-body systems.
In principle one can similarly investigate the lower bound
for general multipartite quantum systems. However as the generalized
Schmidt decomposition of multipartite pure states becomes more
complicated when the number of subsystems increases, the
problem would be more sophisticated.

\bigskip
We gratefully acknowledge the support provided by National Natural
Science Foundation of China Under Grant Nos. 10375038, 90403018,
10271081, 10675086, NKBRPC 2004CB318000 and the Fund of Beijing Municipal
Education Commission KM200510028021. We thank Zhi-Xi Wang for
valuable discussions.

\end{document}